\documentclass{article}
\usepackage[utf8]{inputenc}
\usepackage{lineno}
\usepackage{amsmath,amssymb}
\usepackage{float}
\usepackage[numbers]{natbib}
\usepackage{graphicx}
\usepackage{url}

\newcommand{\etal}{\textit{et al}. }
\newcommand{\ie}{\textit{i}.\textit{e}. }

\newcommand{\me}{\mathrm{e}}
\newcommand*{\vertbar}{\rule[-1ex]{0.5pt}{2.5ex}}
\newcommand*{\horzbar}{\rule[.5ex]{2.5ex}{0.5pt}}

\title{Forecasting Short-term Dynamics of Fair-Weather Cumuli using Dynamic Mode Decomposition}
\author{Jeff Manning and Ross Baldick}
\date{%
	Department of Electrical and Computer Engineering,\\
	University of Texas, Austin, Texas USA}

\begin{document}

\maketitle

\begin{abstract}
	Application of Dynamic Mode Decomposition to clear-sky index forecasting of shadowing effects of convective fair-weather cumulus clouds is presented. Cloud dynamics are captured by sequences of visible-light photographic video frames. This method can be more easily applied to the modeling of cloud evolution than traditional fluid-based methods, and can enhance existing frozen-cloud advection methods. Its use is demonstrated for an actual fair-weather cumulus cloud image sequence and compared to an advection-only forecast. It is concluded that the method shows promise for very short-term clear-sky index forecasting for up to seven minute horizons.
	
\end{abstract}

\section{Introduction}
\subsection{Variabilities caused by Fair-Weather Cumuli}
\begin{figure}
	\centerline{\includegraphics[width=\textwidth]{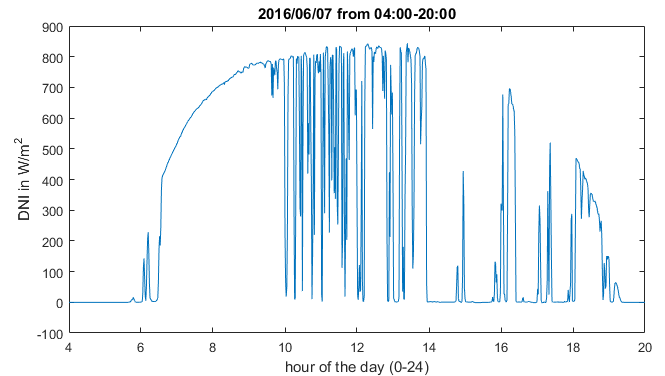}}
	\caption[Direct Normal Irradiance (DNI) vs. time of day for June 7, 2016 at the University of Texas Rio Grande Valley campus.] {Direct Normal irradiance vs. time of day for June 7, 2016 at the University of Texas Rio Grande Valley campus. The early part of the day shows almost no high-frequency variation, but high-frequency variations persist from mid-morning until almost sunset. (Data from \cite{UTPA_Data}).}
	\label{fig:allday_insolation} 
\end{figure}

Variability of renewable energy generation poses challenges to operators of the electric grid. Solar photovoltaic (PV) power is an excellent source of low-cost energy. However, the radiant energy falling on a PV site can drop as much as 70\% from its clear-sky value when clouds pass between the sun and the site \cite{NERC_2009}. Uniquely challenging fluctuations are caused by fields of \textit{cumulus humilis} or ``fair-weather'' cumulus clouds \cite{Pedro_2015}. Such clouds can cause numerous, deep swings in PV generation output that persist for hours, as shown in Figure \ref{fig:allday_insolation}. Fluctuations \cite{Barbieri_2017} and generation ramp rates caused by sharp-edged cumuli are also more severe than others \cite{Jewell_1987}.

Fair-weather cumuli are produced by radiative solar heating of the surface, resulting in a turbulent convective daytime atmospheric boundary layer (ABL). With sufficient air moisture, small, opaque clouds form at the top of the ABL, usually 1-2 km above the surface \cite{WallaceHobbs_2006}. As an example, fields of fair-weather cumuli commonly span areas of  hundreds of square kilometers on summer afternoons in the central Texas region of the U.S., advected by a moderate and usually uniform wind at altitude.

\subsection{Effects of Variability on System Resources}
Heavy PV generation on a single distribution feeder may challenge voltage magnitude regulation in the presence of deep and frequent local swings in active power production. With future advances in behind-the-meter technology, residential inverters and home energy management systems (EMS) will need the ability to optimize the delivery of energy parcels not only to the grid and fixed loads, but also to discretionary loads such as battery energy storage systems (BESS), HVAC resources, and other demand-response-capable equipment.

Battery lifetimes are sensitive to their charge/discharge profiles, and air conditioning compressors cannot be arbitrarily switched on and off under high coolant head pressure. These examples suggest that EMS and BESS resources could better manage real-time energy flows and local power quality from knowledge not only of the current instantaneous irradiance, but also its estimated trajectory over the near future \cite{Kundur_1993}.

\subsection{Solar Irradiation Forecasts}
Solar PV forecasting has historically focused on forecast horizons in the day-ahead to year-ahead range \cite{Kleissl_2013}. However, the large, abrupt, and frequent power output swings caused by shallow cumuli motivate efforts to provide some advance warning of ramp events to inverters, especially at smaller PV plants in distribution systems. 

Many factors can affect the power output of a PV panel, including the angle of the panel relative to the sun, absorbing or scattering particles in the air, and the temperature of the panel. The proposed methods focus only on modeling irradiance variability due to shallow convective cumulus clouds.

Recent work in intra-hour solar forecasting uses whole-sky imaging with cameras, since satellite imagery is still insufficient for spatial resolutions finer than approximately one kilometer. Chow, \etal \cite{Chow_Belongie_Kleissl_2013} defined a confidence metric that tags stable clouds (\ie clouds with essentially constant shapes) with high confidence, and unstable, or evolving clouds with low confidence, since a prediction based on advection or translation of the cloud to an estimated future position produces a poor forecast if the cloud's shape is evolving during the forecast interval. Moncada, \etal \cite{Moncada_2018} employed optical flow techniques for computing cloud motion vectors to predict future clouds as advected from their observed positions, and deep learning methods to map images to irradiance. Kurtz, \etal \cite{Kurtz_2017} found that frozen-cloud advection techniques decline in forecast accuracy for horizons beyond a few minutes, and states, ``further improvements would require a more dynamic model for cloud development.'' 

In the following sections, we introduce advanced dimension-reduction methods to make clear-sky index (CSI) forecasts from a sky image sequence. These methods are shown to capture the high-dimensional image sequence in a reduced model that captures some of the coherent dynamics of shadow-causing cumuli. The methods described herein are novel because the forecasting state of the art does not at present efficiently model the dynamics of shallow cumulus clouds, while the proposed methods can provide such an ability.

\subsection{Modeling of Atmospheric Phenomena}

Historically, practical modeling of dynamic atmospheric processes has used techniques of computational fluid dynamics (CFD) applied to the incompressible Navier-Stokes partial differential equations or some variant thereof \cite{Anderson_1995}. We developed such a model and found that it could indeed compute the formation and evolution of shallow cumuli due to surface heating and convection. However, such CFD-based methods are computationally expensive, of high dimension (the state vector includes every voxel in the domain), and require that boundary conditions be defined for every variable (three velocity variables, temperature, and water vapor mixing ratio) at every boundary voxel of the domain. Since atmospheric conditions are only approximately known, nearly all of the boundary conditions must be estimated, and the nonlinear nature of the Navier-Stokes convective term makes any such simulation very sensitive to those estimated boundary conditions. Furthermore, widespread turbulence throughout the convective ABL challenges CFD computation. Kurtz, \etal \cite{Kurtz_2017} report similar determinations.

Moreover, a cloud forecast for a particular PV site does not need the vast complement of information provided by a fluid model. We are therefore motivated to seek a reduced-order method in a convenient (photographic) measurement basis, with scalable modeling fidelity, and straightforward initial and boundary conditions.

A reduced-order method, Proper Orthogonal Decomposition (POD), has been used for several decades to analyze fluid flows \cite{Holmes_1996}. Dynamic Mode Decomposition (DMD), the method chiefly employed in this paper, is based on POD but produces closed-form temporal components.  Irradiance signals composed of such temporal components are easy to evaluate at future times to provide a forecast. 

DMD has been applied to many problems in fluid dynamics, partly for the above reasons of data parsimony and computational issues. Some fluid applications of DMD are listed in Table \ref{table:dmdfluidapps}. All of the listed applications are fluid problems that exhibit nonlinear dynamics, and for which the physical model is also represented by a form of the incompressible Navier-Stokes equations. DMD provides a way to model the dynamics in comparatively low dimension and represent the temporal modes with complex exponential combinations of growth, decay, and sinusoidal oscillations.

\begin{table}[ht]
	\caption{DMD Fluid Applications} 
	\centering 
	\begin{tabular}{c c l} 
		\hline\hline 
		Author & Application & Citation\\ [0.5ex] 
		\hline 
		Higham & Shallow water flows & 2017 \cite{Higham_2017} \\ 
		Schmid & Cavity flow, Helium jet, Membrane wake & 2010 \cite{Schmid_2010},\cite{Schmid_2010a} \\ 
		Rowley & Jet in a crossflow & 2009 \cite{Rowley_2009} \\ 
		Tu & Bluff-body wake & 2014 \cite{Tu_2014} \\
		Jovanovic & Two-dimensional channel flow & 2014 \cite{Jovanovic_2014} \\
		Zhang & Flow behind a cylinder & 2014 \cite{Zhang_2014} \\ [1ex] 
		\hline 
	\end{tabular}
	\label{table:dmdfluidapps} 
\end{table}
The problem of modeling cumulus cloud dynamics is similar to the above fluid examples in that, in each case, the dimension of the system state is very large, nonlinear dynamics are too prevalent to ignore, but much of the energy can be explained by a comparatively small set of coherent features. DMD has provided novel insights for the above research targets and is similarly attractive for modeling of cumulus cloud dynamics.

The rest of the paper is organized as follows: Section \ref{methods} discusses the method and its practical application to an actual cloud image sequence. Section \ref{results} describes results, comparing the method to a simple frozen-cloud optical flow approach, and Section \ref{conclusions} offers concluding remarks.

\section{Methods}
\label{methods}

This section describes the proposed forecasting method. Section \ref{ssPOD} briefly defines POD and how it can be used to remove the solar disk from an entire sequence of images in one operation. Section \ref{ssDMD} defines DMD and discusses practical aspects of its application. Section \ref{ssDMDCloud} describes implementation of the proposed method using an actual cloud image sequence as an example. 

\subsection{Proper Orthogonal Decomposition}
\label{ssPOD}
\subsubsection{Definition}
\label{ssPODDef}
Consider a linear system with an $N$-dimensional state vector $\mathbf{x}$ which varies over time. For grayscale images, this means the pixel columns of each image are stacked end-to-end to form a single-column (``flattened'') state vector. If $M$ states are uniformly sampled in time, say every $\Delta t$ seconds, and $x_{ik}$ is defined as the $i$-th state element sampled at the $k$-th sample time, an ($N \times M$) matrix can be constructed to represent the time-sampled evolution of the system:

\begin{displaymath}
\mathbf{X} = \begin{bmatrix}	x_{11}&x_{12}& ... &x_{1M}\\
x_{21}&x_{22}& ... &x_{2M}\\
\vdots    & \vdots   & \ddots & \vdots\\
x_{N1}&x_{N2}& ... &x_{NM}\\
\end{bmatrix}.
\end{displaymath}

The singular value decomposition (SVD) $\mathbf{X} = \mathbf{U} \mathbf{\Sigma} \mathbf{V^*}$ (where $^*$ means complex conjugate transpose, $\mathbf{U} \in \mathcal{R}^{N\times N}$ and $\mathbf{V}\in \mathcal{R}^{M\times M}$ are unitary, and $\mathbf{\Sigma}\in \mathcal{R}^{N\times M}$, diagonal with nonnegative entries $\sigma_k$, ordered from largest to smallest magnitude) separates the time dynamics which are captured by the sampled states into a set of \textit{spatial} modes embodied in the columns of $\mathbf{U}$ and \textit{temporal} modes given by the columns of $\mathbf{V}$. The SVD, when computed on a matrix so constructed, is the POD of $\mathbf{X}$ \cite{Higham_2017}. 

If the state elements $x_{ik}$ are the grayscale pixel intensities of the pixels of an $H$ (height) $\times W$ (width) image, and flattened such that the image pixel columns are stacked end-to-end to form an $N \times 1$ vector (where $N=HW$), the POD exposes a set of time series in the columns of $\mathbf{V}$, corresponding to the pixel distributions (the POD spatial modes) over which those time series evolve, represented in the columns of $\mathbf{U}$. For such high-dimensional problems, $N >> M$ is common and is assumed here. The outer product of column $\mathbf{u_k}$ and column $\mathbf{v_k}$, weighted by singular value $\sigma_k$ comprises the $k$-th POD mode. 

The POD of a given data matrix $\mathbf{X}$ can be written as:

\begin{align}
\label{dmdouter}
\mathbf{X} &= \begin{bmatrix}	\vertbar & \vertbar & & \vertbar \\ \mathbf{x_1}&\mathbf{x_2}& ... &\mathbf{x_M}\\ \vertbar & \vertbar & & \vertbar \end{bmatrix}\nonumber\\
&= \mathbf{U \Sigma V^*}\nonumber\\
&=  \begin{bmatrix}	\vertbar & \vertbar & & \vertbar \\ \mathbf{u_1}&\mathbf{u_2}& ... &\mathbf{u_M}\\ \vertbar & \vertbar & & \vertbar \end{bmatrix} 
\begin{bmatrix}	\mathbf{\sigma_1}& &\\
& \ddots &\\
& & \mathbf{\sigma_M}\\
& & \end{bmatrix} 
\begin{bmatrix}	\horzbar &\mathbf{\bar{v}_1} & \horzbar \\ \horzbar &\mathbf{\bar{v}_2}& \horzbar \\ &\vdots & \\ \horzbar & \mathbf{\bar{v}_M} & \horzbar\\ \end{bmatrix}\nonumber\\
&= \begin{bmatrix} \vertbar \\ \mathbf{u_1} \\ \vertbar \\ \end{bmatrix}  \begin{bmatrix} \mathbf{\sigma_1} \end{bmatrix} \begin{bmatrix} \horzbar & \mathbf{\bar{v}_1} & \horzbar \end{bmatrix} + \cdots + \begin{bmatrix} \vertbar \\ \mathbf{u_M} \\ \vertbar \\ \end{bmatrix}  \begin{bmatrix} \mathbf{\sigma_M} \end{bmatrix} \begin{bmatrix} \horzbar & \mathbf{\bar{v}_M} & \horzbar \end{bmatrix},
\end{align}
where $\mathbf{u_k}$ and $\mathbf{v_k}$, $k=1...M$ are the respective columns of $\mathbf{U}$ and $\mathbf{V}$ with overbar meaning complex conjugate; $\mathbf{\sigma_k}$, $k=1...M$ are the diagonal elements of $\mathbf{\Sigma}$; $\mathbf{x_k}$, $k=1...M$ are the flattened images that comprise the columns of $\mathbf{X}$, and each term in (\ref{dmdouter}) is a POD mode.

\begin{figure}[h!]
	\centerline{\includegraphics[width=\textwidth]{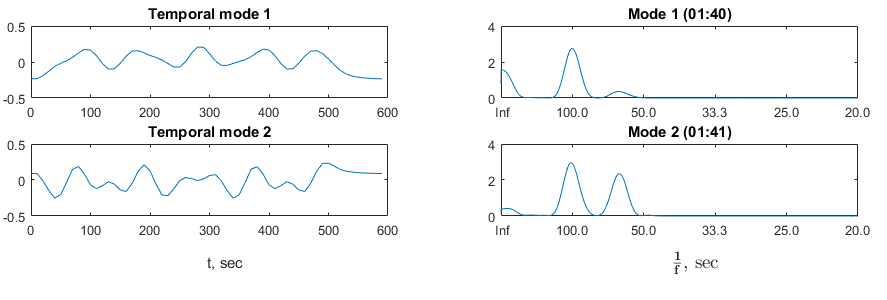}}
	\caption[Example POD Temporal Modes.]{Example POD Temporal Modes. Each mode consists of the entries of one column of the $\mathbf{V} \in \mathcal{R}^{60 \times 60}$ matrix. $\Delta t$ = 10 seconds, so $M\Delta t = 600$ seconds. Left: The actual time series, exactly as contained in columns $\mathbf{v_1}$ and $\mathbf{v_2}$. Right: The magnitude of the FFT of the time series.}
	\label{pod_temporal_modes} 
\end{figure}

Figure \ref{pod_temporal_modes} shows two example POD temporal modes with their respective Fast Fourier Transform (FFT) magnitude spectra. Since each time series may contain multiple spectral components, future temporal mode values cannot be directly computed for $t>M\Delta t$. In contrast, as will be discussed in Section \ref{ssDMD}, DMD produces temporal modes that are specific complex exponentials \cite{Jovanovic_2014}, and can therefore be computed directly at future times, which makes DMD preferable over POD for forecasting.

\subsubsection{Direct use of POD to remove the solar disk}
\label{ssPODSR}

Sky images often contain increased glare in the area of the solar disk, making resolution of clouds difficult near the sun. Although POD is not convenient in general for forecasting, it can be useful for removing the solar disk from the sequence. In particular, because the first POD mode contains the temporally constant component of the sequence, its removal also removes the bright solar disk from all image frames in one operation. We found, however, that complete omission of the first POD mode resulted in decreased image contrast, so the proposed method uses the first mode to locate the solar disk, then the spatial remainder of the first mode is added back to the sequence. 

Once the images have been rotated and cropped so that only the upwind portion remains, each image column represents a set of pixels that are approximately the same upwind distance from the sun. The glare of the sun outside the solar disk can be found and removed by creating a row vector $\mathbf{g}$ whose elements $g_j, j=1, \dots w$ are the minimum pixel value in each image column, smoothing the vector, and subtracting those values from the respective columns of each image. Glare from the sun is thereby effectively removed without affecting resolution of clouds.

\subsection{Dynamic Mode Decomposition} 
\label{ssDMD}
\subsubsection{Definition}
\label{ssDMDDef}

DMD is a method that can approximately model cloud evolution dynamics without the burden of high dimension and the computational cost of a three-dimensional fluid model. Schmid \cite{Schmid_2010} defines DMD as follows: Given a system with $k$th state $\mathbf{x_k} \in \mathcal{R}^{N}$, measure the state over $M-1$ time intervals separated by sampling time $\Delta t$ and produce a $space \times time$ matrix $\mathbf{X} \in \mathcal{R}^{N \times M}$, whose columns are the successive state vectors $\mathbf{x_k}$. Further, define $\mathbf{X_1}, \mathbf{X_2} \in  \mathcal{R}^{N \times (M-1)}$ with $\mathbf{X_1} = [\mathbf{x_1...x_{M-1}}]$ and $\mathbf{X_2} = [\mathbf{x_2...x_{M}}]$. Hence, $\mathbf{X_2}$ is $\Delta t$ forward shifted, and the same size as $\mathbf{X_1}$. Under the assumption that the time evolution of the system is sufficiently modeled as a linear first-order ODE, we may then say that column $\mathbf{x_{k+1}} = \mathbf{Ax_k}$ thus $\mathbf{X_2 = AX_1}$ for some matrix $\mathbf{A} \in \mathcal{C}^{N \times N}$. Let $\mathbf{U \Sigma V^* = X_1}$ be the SVD of $\mathbf{X_1}$.  $\mathbf{U} \in \mathcal{R}^{N\times N}$, $\mathbf{V}\in \mathcal{R}^{(M-1)\times (M-1)}$, and $\mathbf{\Sigma}\in \mathcal{R}^{N\times (M-1)}$ diagonal with nonnegative entries $\sigma_{k}, k = 1, \dots, M-1$ on its diagonal. The columns of $\mathbf{U}$ are the POD spatial modes of $\mathbf{X_1}$ and describe how the POD temporal components (time series exposed in the columns of $\mathbf{V}$) are applied spatially \cite{Kutz_2016}. Under the assumption that the system is governed by a linear first-order ODE, we can write:

\begin{align*}
\mathbf{X_2} &= \mathbf{AX_1} \\
&= \mathbf{AU \Sigma V^*}. 
\end{align*}

Define:

\begin{equation*}
\mathbf{\tilde{A}} := \mathbf{U^*AU} = \mathbf{U^*X_2V\Sigma^{-1}}.
\end{equation*}
The matrix $\mathbf{\tilde{A}} \in \mathcal{C}^{N \times N}$ is the projection of $\mathbf{A}$ onto the POD modes of $\mathbf{U}$. $\mathbf{\tilde{A}}$ is related  to $\mathbf{A}$ by a similarity transformation.  The eigendecomposition of $\mathbf{\tilde{A}}$ is given by:

\begin{equation*}
\mathbf{W M W^{-1}} = \mathbf{\tilde{A}},
\end{equation*}
where $\mathbf{W}$ is the unitary matrix of eigenvectors of $\mathbf{\tilde{A}}$ and $\mathbf{M}$ is the diagonal matrix of corresponding eigenvalues $\mu_{i}, i=1,\dots,N$. If $\mathbf{A}$ and therefore $\mathbf{\tilde{A}}$ are limited to rank less than or equal to $r$, then an $r \times r$ matrix that captures their dynamics can be defined by:

\begin{equation*}
\label{dmddef2r}
\mathbf{\tilde{A}_r} := \mathbf{U_r^*AU_r} = \mathbf{U_r^*X_2V_r\Sigma^{-1}_r},
\end{equation*}
where $\mathbf{U_r}$ and $\mathbf{V_r}$ are comprised of the first $r$ columns of $\mathbf{U}$ and $\mathbf{V}$ respectively, and $\mathbf{\Sigma_r}$ is the top-left $r \times r$ block of $\mathbf{\Sigma}$. The eigendecomposition of $\mathbf{\tilde{A}_r}$ is:

\begin{equation*}
\mathbf{\tilde{A}_r} := \mathbf{W_r \Lambda W_r^{-1}}.
\end{equation*}

The individual eigenvalues of $\mathbf{\tilde{A}_r}$ are the diagonal entries $\lambda_{i}, i=1,\dots,r$ of $\mathbf{\Lambda}$. If $rank(\mathbf{A}) = rank(\mathbf{\tilde{A}}) \le r$, then the  $\lambda_{i}$ will be equal to the $\mu_{i}$ associated with the highest-energy POD modes of $\mathbf{A}$. Moreover, they are found without performing an eigendecomposition of a $N \times N$ matrix, which would be computationally expensive and possibly intractable. 

Once the $\lambda_{i}$ are found, the actual complex arguments $\omega_{i}$ of the continuous-time differential equation whose discrete analog is represented by the operator $\mathbf{\tilde{A}_r}$ are computed as:

\begin{equation}
\label{dmddef4a}
\Omega = diag([\omega_1,\dots,\omega_r])\text{, where} \quad \omega_i =  \frac{1}{\Delta t}log(\mathbf{\lambda_i}), i = 1,\dots,r.
\end{equation}

The modes of $\mathbf{\tilde{A}_r}$ exist in the reduced $r$-dimensional space. The DMD modes (in the original $N$-dimensional space) are recovered in the columns of:

\begin{equation*}
\mathbf{\Phi = X_2 V_r \Sigma_r^{-1} W_r}.
\end{equation*}
$\mathbf{\Phi} \in \mathcal{C}^{N \times r}$ and its columns are how the time dynamics given by the $\omega_i$ are distributed spatially in the original images. A low-rank approximation of the original $\mathbf{X}$ is then computed as:

\begin{equation}
\label{dmddef5}
\mathbf{X_{dmd}} = \mathbf{\Phi} \me^{\mathbf{\Omega} t} \mathbf{b}.
\end{equation}
where $\mathbf{b} \in \mathcal{C}^{r}$ is found from the initial condition $\mathbf{x_0} \in \mathcal{R}^{N}$, \ie setting $t=0$ in (\ref{dmddef5}):

\begin{equation*}
\mathbf{\Phi b = x_0} \implies \mathbf{b = \Phi^{+} x_0}. 
\end{equation*}
The notation $^+$ means Moore-Penrose pseudoinverse. The initial condition $\mathbf{x_0}$ is simply the first image of the sequence, flattened into a column vector. No other initial or boundary conditions are needed.

The efficacy of DMD is to discover a relatively small set of coherent time dynamics and their spatial placement from a set of high-dimensional data, with no need for a high-dimensional eigendecomposition. For example, the columns of $\mathbf{X}$ may come from a sequence of $100 \times 100$ pixel images, meaning that $\mathbf{A} \in \mathcal{R}^{10000 \times 10000}$. If only a small number, say ten significant temporal modes are present in the captured dynamics, we may thus reduce the dimension to $r = 10$ and operate in a 10-dimensional space with $\mathbf{\tilde{A}_r} \in \mathcal{R}^{10 \times 10}$.

Moreover, forward prediction in time is facilitated by DMD, since to evaluate $\mathbf{X_{dmd}}$ over a future interval $[t_{f1}, t_{f2}]$, (\ref{dmddef5}) may be directly computed for those $t \in [t_{f1}, t_{f2}]$.

\subsubsection{Augmentation of State Vectors}
\label{ssDMDAug}
Much of the power of DMD comes from the assumption that the time dynamics are sufficiently modeled by a linear first-order difference equation, i.e. $\mathbf{X_2 = AX_1}$. However, many systems produce an $\mathbf{X}$ with numerous periodic components. Such a periodic constituent would reveal itself (with real-valued data) as a pair of conjugate imaginary eigenvalues \cite{Tu_2014}. However, there is no first-order difference equation that can operate on the reals to produce a periodic time series. Consider the system with only one state element $x_k$ at time step $k$:
\begin{equation}
\label{dmdaug1}
x_k = a x_{k-1}, a \in \mathcal{R};
\end{equation}
which produces:
\begin{displaymath}
\mathbf{X} =  \begin{smallmatrix}[x_0 & x_1 & \dots & x_M] \end{smallmatrix}  = \begin{smallmatrix}[ x_0 & a x_0 & a^2 x_0  & \dots & a^{M}x_0] \end{smallmatrix}\text{,}
\end{displaymath}
so for real-valued $a$, $x$ can only grow or decay exponentially from its initial value. If $x$ is allowed to have complex values, exponentiation changes from being a simple successive scaling operation into a simultaneous scaling \textit{and rotation} operation about the origin of the complex plane. Successive $x_k$ then have both magnitude and angle and are \textit{inherently} periodic, since the complex value of $x$ is spinning around the complex origin, in addition to exponentially growing or decaying, in the case that the real part of the exponent is not zero.

The augmented DMD allows $\mathbf{A}$ to represent a higher-order differential equation by lifting the state vector to a higher dimension. In practical terms, \cite{Kutz_2016} describes the augmented DMD as an application of the ``shift-stacking'' operation employed by \cite{Tu_2014} to increase the rank of the data matrix $\mathbf{X}$ so that the time-step transformation $\mathbf{A}$ can have enough eigenvalues to describe the dynamics of $\mathbf{X}$. We offer the following alternative view.  Consider the second-order difference equation:

\begin{equation}
\label{dmdaug2}
x_{k} = a x_{k-2} + b x_{k-1}.
\end{equation}
Converting the second-order equation in $\mathcal{R}$ to a first-order equation in $\mathcal{R}^2$, we may define $y_k = x_{k-1}$ so that $x$ becomes two-dimensional. The equation

\begin{equation}
\hat{x}_k = \begin{bmatrix}y_k\\ x_k \end{bmatrix} = \begin{bmatrix} 0 & 1\\ a & b \end{bmatrix} \begin{bmatrix}y_{k-1}\\ x_{k-1} \end{bmatrix} = A\hat{x}_{k-1}
\end{equation}
is thus a first-order difference equation for $\hat{x}_{k} \in \mathcal{R}^2$ and is equivalent to (\ref{dmdaug2}), which is a second-order difference equation for $x_{k} \in \mathcal{R}$. 

When implementing the augmented DMD, we similarly begin with $\mathbf{X_1} = \begin{smallmatrix}[ \mathbf{x_1} & \mathbf{x_2} & \mathbf{x_3} & \dots & ] \end{smallmatrix}$ and $\mathbf{X_2} = \begin{smallmatrix}[ \mathbf{x_2} & \mathbf{x_3} & \mathbf{x_4} & \dots & ] \end{smallmatrix}$, but for column $\mathbf{x_k}$ made up of real-valued elements, there is no $\mathbf{A}$ such that $\mathbf{X_2 = A X_1}$ when 2nd or higher-order dynamics are present in the data. However, the augmented system:

\begin{equation}
\label{dmdaug}
\mathbf{\hat{X}_2} = \begin{bmatrix}\mathbf{x_2}&\mathbf{x_3}&\mathbf{x_4} ...\\\mathbf{x_3}&\mathbf{x_4}&\mathbf{x_5} ... \end{bmatrix} = \begin{bmatrix} \mathbf{0} & \mathbf{1}\\ \mathbf{a} & \mathbf{b} \end{bmatrix}\begin{bmatrix}\mathbf{x_1}&\mathbf{x_2}&\mathbf{x_3} ...\\\mathbf{x_2}&\mathbf{x_3}&\mathbf{x_4} ...\end{bmatrix} = \mathbf{A} \mathbf{\hat{X}_1}
\end{equation}
faithfully reproduces the 2nd-order relationship. Thus the similarity between reducing a 2nd-order difference equation to a set of first order difference equations and the augmented DMD model is shown. Although the proposed method uses only one level of augmentation as described above, more levels may be applied to achieve linear consistency of $\mathbf{X_1}$ and $\mathbf{X_2}$, which is required by \cite{Tu_2014}. Augmentation is used in Section \ref{ssDMDCloud}.

\subsubsection{Rank Truncation}
\label{ssDMDRank}

Measurements of natural phenomena are not likely to have an absolute rank limit. Turbulent dynamics are commonly expected to have a smooth spectral decay all the way down to wavenumbers at the Kolmogorov microscale \cite{Holmes_1996}, which is much smaller than the effective pixel resolution of a sky image. Consequently, any reduced order model must truncate the rank to some computationally practical value of $r$, meaning that some (high wavenumber) coherent energy must instead be modeled by other (low wavenumber) modes. Truncation will result in eigenvalues of $\mathbf{\tilde{A}_r}$ that \textit{approximate} some of the eigenvalues of $\mathbf{A}$. However, because they are projected onto the first (and most energetic) principal components of the data, the reduced-order eigenvalues produce an $\mathbf{X_{dmd}}$ that minimizes the L2-norm of the reconstruction error. 

Efforts to decompose the dynamics of observed clouds might be useful in identifying specific coherent dynamics, for example, representing specific vortical circulations. If energy is discarded by order truncation, however, the DMD eigenvalues of $\mathbf{\tilde{A}_r}$ will only be \textit{bounded} by the true eigenvalues of $\mathbf{A}$, as suggested by the Cauchy Interlacing Property \cite{Antoulas_2005}. It is consequently important to distinguish between the best-fit reduced-order model and the specific values of particular eigenvalues, and not to expect, for example, a particular complex eigenvalue to represent the true rotation period of a specific physical vortical structure in the cloud observations. 

\subsection{DMD of a Cloud Image Sequence}
\label{ssDMDCloud}

In this section, the method is applied to an image sequence that contains a decaying cloud, to demonstrate how the inclusion of dynamical information provided by DMD can offer improved forecasts over a frozen-cloud advection model. This example is only intended to demonstrate that the proposed method can be used to improve existing forecasting techniques by accounting for some degree of temporal cloud evolution in the forecast model.

In the following analysis, grayscale pixel value $C$ is used as a measure of cloudiness. CSI $K$ can be estimated as $K = 1 - C$. Given a uniform horizontal wind, it is straightforward to project a set of cloud pixels forward in time, and future $K$ is estimated simply as the sum of the cloudiness of all future pixels within the solar disk, divided by total pixel area of the solar disk.  

It is acknowledged that the above measures are too simple for actual irradiance forecasts. Moreover, an effective mapping between $K$ and the true DNI or GHI falling on a PV array is not linear and is affected by phenomena not addressed herein. Our purpose is however to identify clouds in the camera frame and to make a useful measure of their evolution. The proposed method could potentially be integrated with other useful forecasting techniques, with the assumption that many other phenomena must be modeled in the process. Modeling of short-term cloud dynamics is claimed to be only one contribution to the greater solution. 

\begin{figure}[h!]
	\centerline{\includegraphics[width=\textwidth]{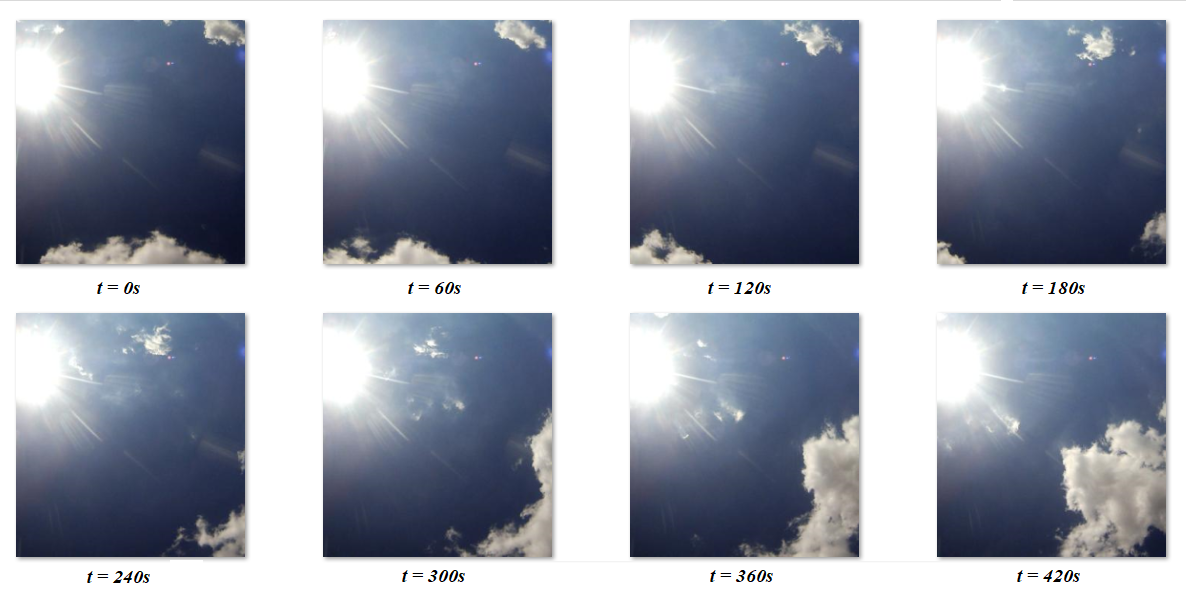}}
	\caption[Sample Images from Observed Sequence.]{Sample Images from Observed Sequence. $M = 210$, $\Delta t$ = 2s, so $M\Delta t = 420s$ or 7 minutes. A cloud enters the frame at top right and advects towards the sun as it decays. (Only every thirtieth image is shown.)}
	\label{dmd1_obs_tiles} 
\end{figure}

Sky images were extracted from a video collected on May 5, 2019 in the Austin, Texas area. Figure \ref{dmd1_obs_tiles} shows snapshots at sixty-second intervals from the seven-minute sequence of 210 images with $\Delta t = 2s$ to which the method was applied.

The lowest-order (first) POD mode (the first term in (\ref{dmdouter})) contains the temporally constant component of the sequence, which includes the solar disk. The solar disk is located within the frame by thresholding the first POD mode, then adding the portion of the first mode not inside the solar disk back to the sequence. The location of the sun is assumed to be constant over the $M \Delta t$ s interval. This process was described in Section \ref{ssPODSR}.

\begin{figure}
	\centerline{\includegraphics[width=\textwidth]{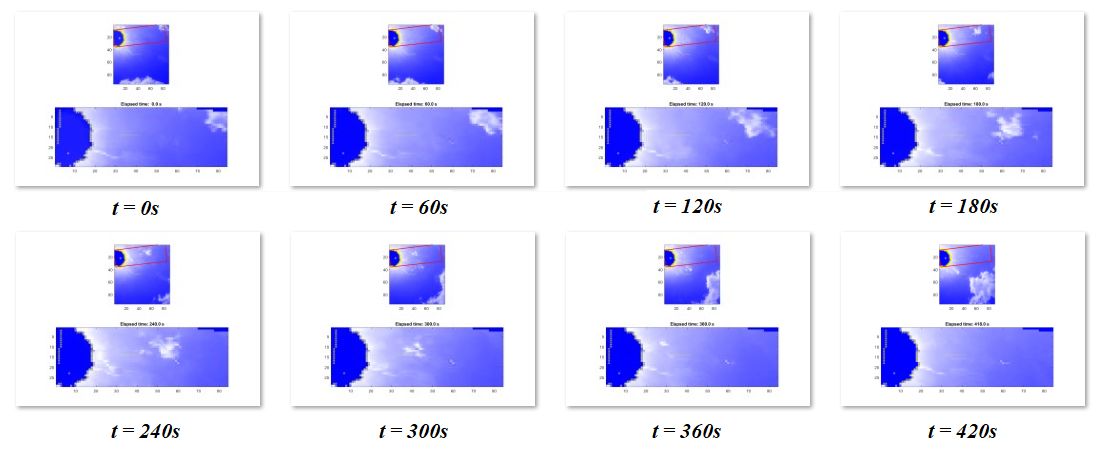}}
	\caption[Images after solar disk removal, rotation and cropping.]{Images after solar disk removal, rotation and cropping. The red box in the top image of each pair shows the solar-upwind region.}
	\label{dmd1_obs_tilesr} 
\end{figure}

The uniform wind is estimated by Horn \& Schunck Optical Flow \cite{Horn_1981}, and the sequence is rotated so motion is from right to left, and only the rectangle of the frame upwind of the sun is retained. Images of the resulting sequence are shown in Figure \ref{dmd1_obs_tilesr}.

Once all images in the sequence are oriented with the solar disk on the left hand side and uniform advection from right to left, forecasts are computed by the following process:
\begin{enumerate}
	\item[1.] Let the set of time sequenced images be $\{\mathbf{I}_k, k=1...M\}$. The image dimension is $H \times W$. At time index k, choose an ordered subset $\mathcal{I}_k = \{\mathbf{I}_k, \dots \mathbf{I}_{k+M_m-1}\}$, with $M_m > r+1$ where $r$ is the desired DMD order ($r=3$ in this case). $M_m$ is the number of  images in the subsequence, thus the number of columns of the data matrix $\mathbf{X}$. Each $\mathbf{I}_k$ is a $H \times W$ image that, when flattened, becomes the $k$th column of the data matrix $\mathbf{X}$.\\ 
	
	\item[2.] Compute the POD of $\mathbf{X}$ built from the images of $\mathcal{I}_k$. For a cloud advecting across the frame, the first POD mode will cover the horizontal span of advection during the period of the subsequence. The spatial extent of this mode identifies boundaries of an appropriate $h \times w$ image inset that contains the cloud's excursion in the frame over the period. If several cloud regions are concurrently moving across the frame, several separate non-overlapping insets may be chosen. In choosing inset segmentation, it is important that a given inset has its own dynamics- that is, two different clouds at different stages of lifecycle should have separate insets to retain their unique dynamics.\\
	
	\item[3.] Compute a third-order ($r=3$) DMD of the sequence of insets. The data matrices $\mathbf{X_1}$ and $\mathbf{X_2}$ each have $M_m-2$ columns since 2nd-order dynamics require 1-deep augmentation, as described in Section \ref{ssDMDAug}. Both matrices then have $2N = 2hw$ rows. The choice of third order is because a minimal model is desired for this test of the method, but we wish to accommodate both a single real-valued exponential component (growth or decay) and an oscillatory component that can represent the typical growth-followed-by-decay pattern of a shallow cumulus cloud, as \{$\sin t$: $0\le t \le \pi$\} does. Recall that an oscillatory component requires a pair of complex-conjugate eigenvalues.
	
	\item[4.] Evaluate $\mathbf{X_{dmd}}$ as defined by (\ref{dmddef5}), at the future time $t_s$ at which that inset is expected to reach the solar disk. The resulting $\mathbf{X_{dmd}}$ represents the expected cloudiness due to clouds within the frame inset at that time. The future (not yet advected) state of the inset at $t_s$ is constructed by ``un-flattening'' the first $N$ rows of $\mathbf{X_{dmd}}$ to an image of size $h \times w$.
\end{enumerate}
Repeat steps 3-4 for each inset, translate the insets forward in time, and construct a composite future $H \times W$ forecast image.

\begin{figure}[H]
	\centerline{\includegraphics[width=0.5\textwidth]{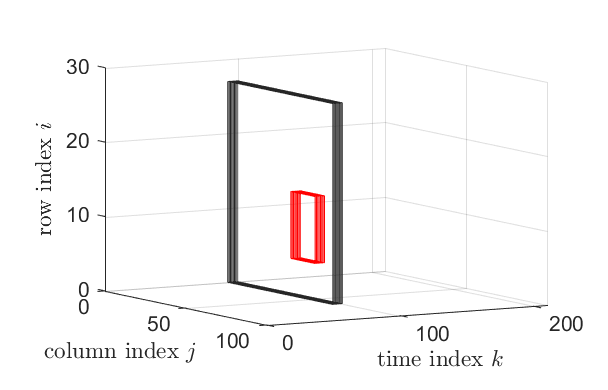}}
	\caption[Inset sequence $\mathcal{I}_{90}$ at $t$=180s.]{Inset sequence $\mathcal{I}_{90} = \{I_{90}, \dots I_{97}\}$ at $t$=180s ($\Delta t = 2s$). $M_m=8$. The inset makes its way through decreasing column indices (right to left) as the cloud advects towards the sun. In a sequence with several clouds, there would be different insets, each modeling the dynamics of the content of its area.}
	\label{dmd_inset_schematic} 
\end{figure}


The above method is performed at every time step to produce a new forecast for a maximum horizon $t_{max} = w \Delta t/u_{avg}$, where $u_{avg}$ is the average advection speed towards the sun in pixels per time step, $\Delta t$ is the length of each time step in seconds, and $w$ is the width of the rotated image frame in pixels.

\section{Results}
\label{results}
For this example, the proposed method (labeled ``DMD'') models cloud dynamics with a simple third-order ($r=3$) DMD, then advects the future estimated cloud forward in time by using the uniform wind velocity estimate found by the optical flow method. Its forecasts are compared with a frozen-cloud advection method (labeled ``Optical flow'') that performs the optical-flow advection on the existing cloud without attempting to capture its temporal evolution.

In the example sequence, a cloud is present and is moving toward the sun. It is however in the decay phase of its life span, and actually dissipates before it reaches the sun. Actual irradiance at the site is therefore ultimately unaffected by the cloud. If a frozen-cloud advection scheme is used to forecast future CSI $K$, it would place the cloud in the vicinity of the solar disk at approximately $t=370$s, and would therefore forecast decreased $K$ at that time. Since the proposed method produces a dynamical model of the cloud, it is able to estimate the future condition of the cloud, then advect the \textit{future} cloud forward in space, yielding a more accurate forecast. In this case, the proposed method finds that the future cloud will dissipate before it reaches the sun and thus forecasts $K$ near 1 as early as $t=30$s, whereas the optical flow forecast continues, in forecasts produced as late as $t=300s$, to expect the cloud to eventually block light from the sun.  

\begin{figure}[H]
	\centerline{\includegraphics[width=\textwidth]{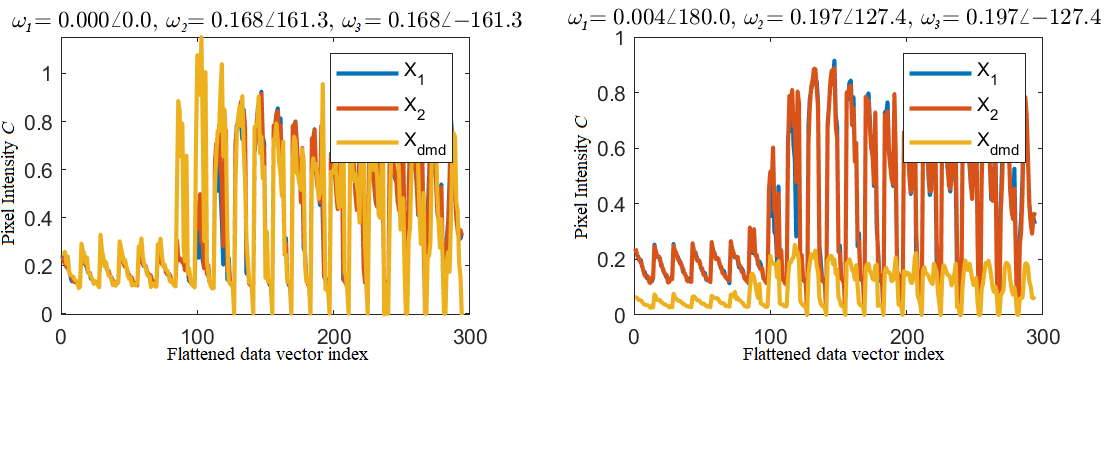}}
	\caption[$\mathbf{X_1}$, $\mathbf{X_2}$, and $\mathbf{X_{dmd}}$ at steps 15 and 16.]{$\mathbf{X_1}$, $\mathbf{X_2}$, and $\mathbf{X_{dmd}}$ at steps 15 (left) and 16 (right). Vertical axis is normalized pixel intensity (values near 1 are cloudy and values near 0 are sunny). Complex exponential arguments $\omega_1, \omega_2, \omega_3$ described in (\ref{dmddef4a}) are listed at the top of each plot. The number of ``humps'' in each plot corresponds to the number of columns in the inset.}
	\label{DMDs_15_16} 
\end{figure}

Figure \ref{DMDs_15_16} shows DMD computation of future clouds. The figure shows details of the data $\mathbf{X_1}$, $\mathbf{X_2}$, and forecast $\mathbf{X_{dmd}}$ at time steps 15 and 16. The method finds a set of three eigenvalues (listed in polar form above each plot) that comprise the best order-3 transformation of $\mathbf{X_1}$ to $\mathbf{X_2}$. If the signal $\mathbf{X_1}$ must decay to form $\mathbf{X_2}$, then the eigenvalues will have magnitudes less than unity and the $\omega_i$ will have negative real parts, as on the right-hand side plot of the figure. Note that at step 15, the forecasted $\mathbf{X_{dmd}}$ is approximately the same magnitude as $\mathbf{X_1}$ and $\mathbf{X_2}$, whereas in time step 16 on the right-hand side, the method forecasts a much-reduced $\mathbf{X_{dmd}}$ primarily because $Re(\omega_1) < 0$. Time step 16 is the first iteration at which the DMD forecast detects dissipation of the cloud.

The entire sequence consists of 210 time steps. The method first forecasts future dissolution of the advecting cloud at time step 16, corresponding to $t=28$s, and continues to forecast $K\approx 1$ at all 194 subsequent frames except frames 25, 26, 31, 32, 39, 40, and 46. Thus the proposed method was able to produce a much more accurate forecast almost five minutes sooner.

\begin{figure}[H]
	\centerline{\includegraphics[width=\textwidth]{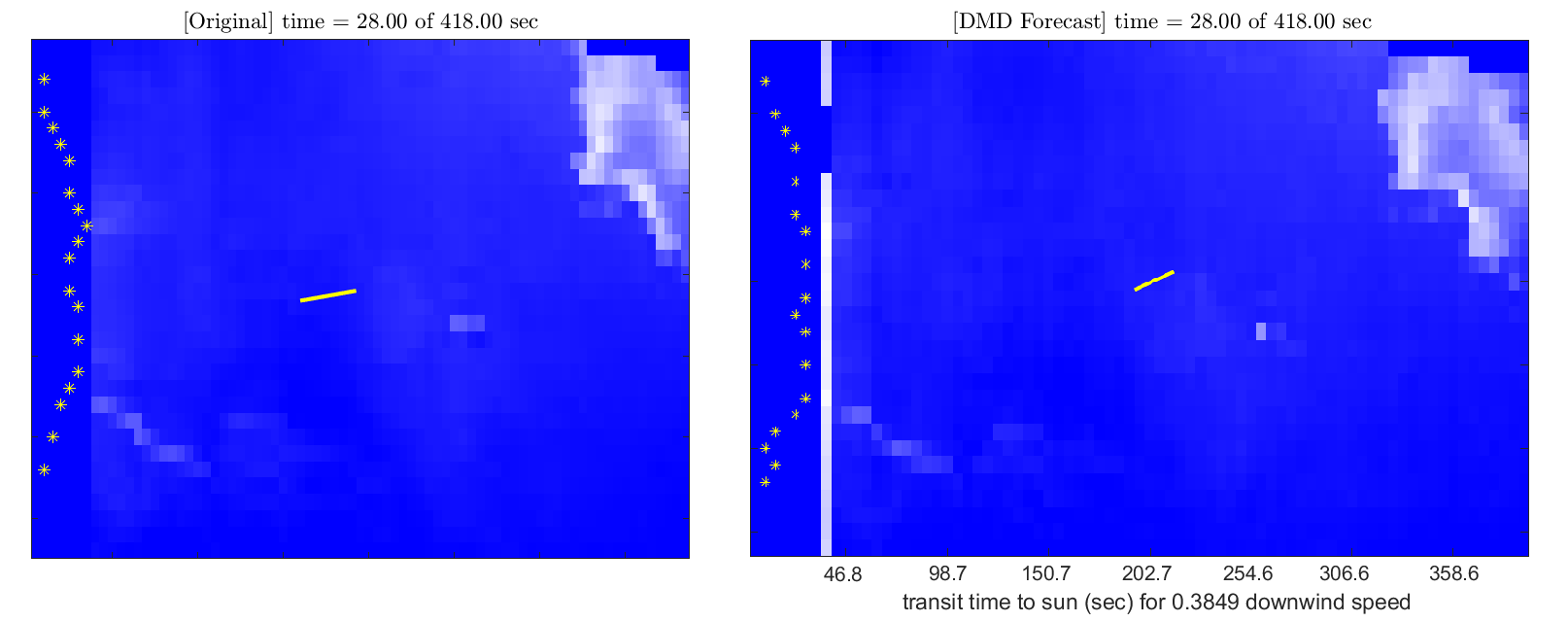}}
	\caption[Original versus future cloud (without advection) at step 15.]{Original versus future cloud (without advection) at step 15 ($t$=28s). At this step, the method has not yet modeled decay of the cloud.}
	\label{orig_forecast_15} 
\end{figure}

\begin{figure}[H]
	\centerline{\includegraphics[width=\textwidth]{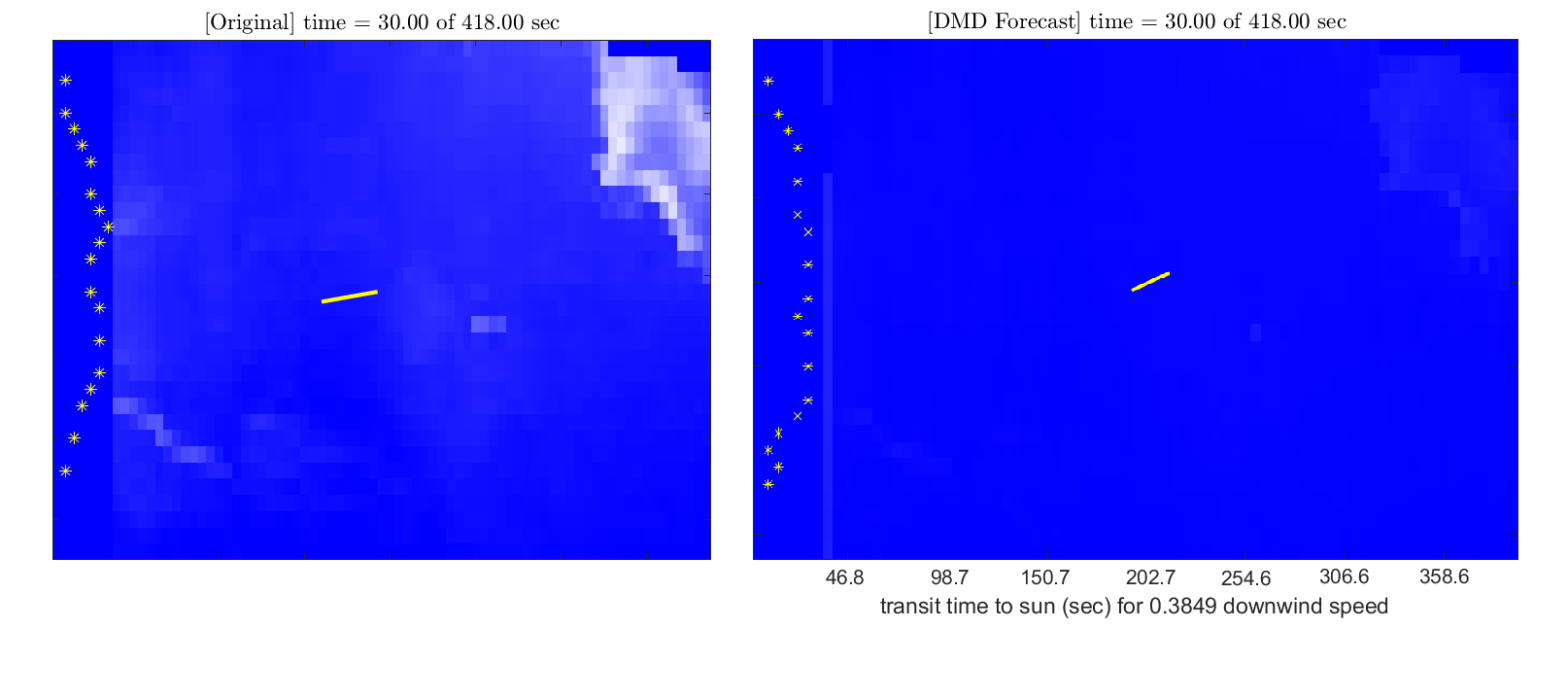}}
	\caption[Original versus future cloud (without advection) at step 16.]{Original versus future cloud (without advection) at step 16 ($t$=30s). This is the first step at which the method models decay of the cloud, forecasting it to have dissipated before reaching the sun, hence there is no visible cloud in the forecast frame.}
	\label{orig_forecast_16} 
\end{figure}

\begin{figure}[H]
	\centerline{\includegraphics[width=\textwidth]{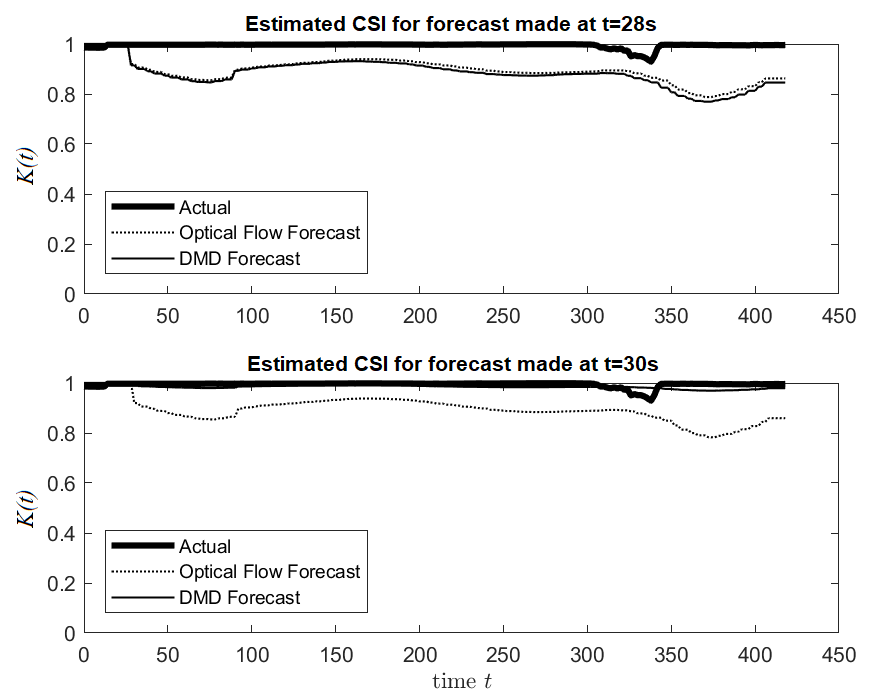}}
	\caption[Actual versus forecast CSI at steps 15 and 16.]{Actual versus forecast CSI at steps 15 (top) and 16 (bottom). At step 15, both methods expect a decreased $K$ at $t=370$s. However, at step 16, the proposed method forecasts that the cloud will have dissipated, and therefore forecasts a high $K$, while the optical flow method continues to expect the cloud to partially block the sun.}
	\label{Forecast_15_16} 
\end{figure}

\subsection{Regarding the irradiance minimum at $t=328$s}

\begin{figure}[H]
	\centerline{\includegraphics[width=\textwidth]{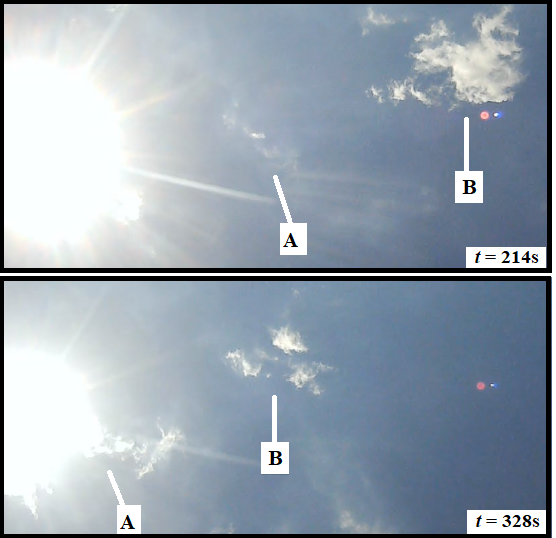}}
	\caption[Growing cloud at $t=214$s (top), and $t=328$s (bottom).]{Growing cloud indicated by marker [A], originating at $t=214$s (top) and partially obscuring the sun at $t=328$s (bottom). Note the smaller apparent solar disk in the bottom image. Marker [B] indicates the cloud modeled in Section \ref{ssDMDCloud}.}
	\label{new_growing_cloud} 
\end{figure}

The only drop in actual CSI in the example sequence occurs during the $t \in [310, 340]$s interval, indicated by the dip in actual CSI in figure \ref{Forecast_15_16}. It is caused by a cloud that originated at approximately $t=214$s and partially obscured the sun at $t=328$s, shown in Figure \ref{new_growing_cloud}. This event is not otherwise addressed herein, but is mentioned so its cause is clear.

\subsection{Computational Performance}
No specific measurements were made in this work for performance gains of DMD over direct eigendecompositions of the $\mathbf{A}$ matrix. However, we did observe that processing demands for direct eigendecompositions were practically prohibitive for all but the smallest of those matrices. For sequences having  $(N,M) \approxeq (10000,100)$, most DMD computations required 3-7 seconds to compute in the MATLAB environment on a modern personal desktop PC. This computation budget is expected to be within the reasonable processing bandwidth of a modern embedded processor that could be cost-effectively deployed in or alongside a PV inverter package.

\section{Conclusions and Future Work}
\label{conclusions}
\subsection{Conclusions}
 In some areas, fields of shallow cumulus clouds are the primary cause of frequent, deep PV generation swings that can last for hours. Such clouds evolve on scales of minutes to tens of minutes, so any forecast more than approximately half of a cloud lifetime into the future must include some model of cloud evolution.
 
 The purpose of this paper is to demonstrate that DMD-modeled dynamics of shallow cumuli can help produce more accurate forecasts over short horizons up to one cloud lifetime. An example is given that demonstrates effective application of the method for a horizon of seven minutes.  We believe that the proposed method can be used to enhance existing forecast models. To our knowledge, existing short-term forecasting techniques do not model fair-weather cumulus cloud dynamics. 

The proposed method identifies individual cloud phenomena in the solar-upwind image frame and models their temporal dynamics as they are advected toward the sun. The example case described in Section \ref{ssDMDCloud} is presented, in which the method identifies a cloud that will dissipate before it reaches the sun, and produces a forecast superior to a frozen-cloud advection forecast. 

Although only a simple case is shown here, we believe that DMD is a powerful technique for modeling evolution of fair-weather cumuli and could be effectively integrated into short-term forecasting mechanisms for such clouds.

Another result demonstrated herein is the use of POD to locate and remove the solar disk and associated glare from the entire image sequence in one operation.

\subsection{Future Work}

DMD is especially useful for the cumulus cloud forecasting problem because it offers a computationally efficient method to produce a forecasting-friendly reduced model. Moreover, model accuracy is easily tuned by parsimonious selection of order $r$. It also offers a best-fit characteristic due to the inherent mode-ordering property of the SVD.

However, DMD is only one method of extracting coherent dynamics from a set of observed data. We foresee a possible role for Independent Component Analysis (ICA) in the decomposition of cloud dynamics, whereby an image sequence might be decomposed into statistically uncorrelated components, rather than the variational ones discovered by POD/DMD.

Recent research has suggested that partial differential equation (PDE) fluid models can aid in cloud forecasting \cite{Yang_2018}. While three-dimensional PDE simulations can be computationally burdensome, there may be promise in the \textit{discovery} of a suitable two-dimensional PDE model that operates in the photographic image spatial domain, using the method of Sparse Identification of Nonlinear Dynamics (SINDy) \cite{Brunton_2015}. SINDy is computationally costly, but is performed once, offline, and discovers a suitable PDE model that optimally fits the data. Once such a model is found, it can be used to perform simulations that may well be within computational capabilities of embedded computing facilities like those in a PV system.

\bibliographystyle{plain}
\bibliography{ms}
\end{document}